# First Demonstration of Field-Deployable Low Latency Hollow-core Cable Capable of Supporting >1000km, 400Gb/s WDM Transmission

A. Saljoghei[1], M. Qiu[2], S. R. Sandoghchi[1], C. Laperle[2], M. Alonso[1], M. Hubbard[2], I. Lang[1], M. Pasandi[2], Y. Chen[1], M. Petrovich[1,3], A. Appleyard[1], A. Boyland[1], L. Hooper[1], T. Bradley[3], G. Jasion[3], H. Sakr[3], E. Numkam Fokoua[3], J. Hayes[3], F. Poletti[3], D. Richardson[1,3], M. Fake[1], M. O'Sullivan[2]

[1]*Lumenisity, Abbey Park Industrial Estate, Romsey, SO51 9DL, UK mike.fake@lumenisity.com*
[2]*Ciena, Ottawa, Ontario, K2K 0L1, Canada mosulliv@ciena.com*
[3]*Optoelectronics Research Centre, University of Southampton, Southampton, S017 1BJ, UK*

**Abstract:** We report order-of-magnitude improvements in performance of field-deployable hollow-core fiber cables evidenced by a 38.4Tb/s (800Gb/s-x-48WDM-channels) 20.5km lab-trial using commercial terminal equipment and the demonstration of 1128km/126km reach in full-fill 400/800Gb/s WDM recirculating-loop experiments. © 2021 The Author(s)

**PDP submitted to but not accepted by OFC 2021 subcommittee D1**

## 1. Introduction

Network transit latency has become a critical issue for an increasing number of time sensitive applications including financial trading, Data Center Interconnect (DCI), 5G and Edge. Hollow-core fiber (HCF) technology offers a unique solution where signals propagate in an air core and travel 50% faster than in standard single mode fiber (SSMF), providing a ~3 µs/km round trip latency benefit [1]. The financial industry, where nano-second latency savings lead to significant commercial advantage, has been an early adopter [2]. More recently, with rapid advancement in hollow-core cable performance, longer reach applications with higher throughput requirements have emerged, such as DCI where HCF solutions offer increased separation between data centers whilst maintaining network latency. Field-deployability, simple installation and backward compatibility with existing network equipment are critical for these applications.

Early HCF work focused on Photonic Bandgap Fiber (PBGF). PBGF data transmission performance is limited by a relatively high loss (~1.7 dB/km the lowest ever reported [3]). Recently, Nested Antiresonant Nodeless Fiber (NANF) designs have emerged [4]. These provide significantly lower loss (~0.28 dB/km reported so far [5]) and have shown evidence of intermodal interference (IMI) at -50 dB/km [6]. Transmission of 61 WDM channels (32 GBd PM-QPSK) through 618 km of NANF has been shown with spooled NANFs in re-circulating loop experiments [6]. However, it is known that cabling a fiber can induce micro- and macro-bend perturbations that affect key properties such as loss, PMD, PDL and IMI. Prior published work on PBGF cables reported ~5 dB/km losses and the transmission of 33x10 Gb/s signals over a 3.1 km distance [7]. Here we demonstrate two NANF cable types (riser for indoor and LT for external use) that support terabit/s capacities and distances over 1000 km. These results represent a major advance in deployable HCF technology and open the door to latency sensitive DCI, 5G and Edge applications. Both NANF Riser and LT cable types have been widely deployed in financial trading applications at 10 Gb/s data rates.

## 2. Field-Deployable Hollow-Core NANF Cable

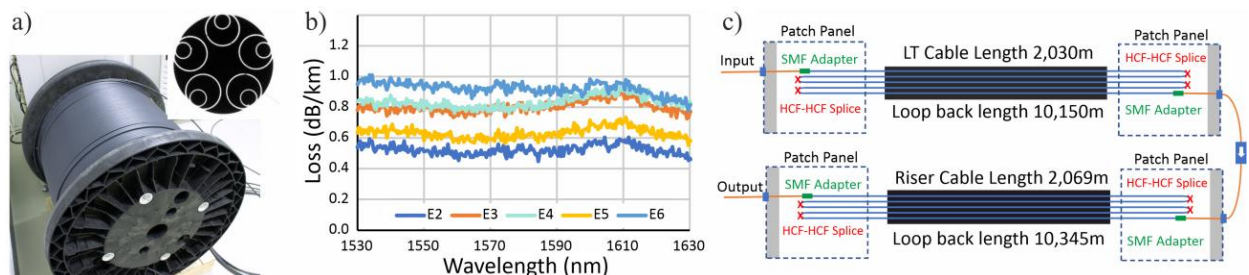

Figure 1: (a) NANF hollow-core cable on a drum as used in these experiments; (b) Loss Spectrum of the five individual 2 km elements in the NANF LT Cable, (c) Configuration of the 20.5 km NANF cable transmission line.

The two field-deployable NANF cables are >2 km long and are wound on two 0.5 m diameter cable drums (see Fig. 1a). The 5 nested-element NANF fiber design (inset) provides an excellent trade-off between loss, micro/macrobend insensitivity and level of IMI experienced in the cable. Each cable contains 5xNANF strands. In Fig. 1b, we show the spectral transmission plot of the 5 NANF strands in the LT cable. The best NANF strand (E2) in the cable has a loss of <0.6 dB/km across the C and L bands. These strands are then spliced together and looped back to create a total cabled NANF fiber transmission link of 20.5 km (see Fig. 1c). The NANF splices were made using a commercial fusion splicer with a proprietary splicing program that ensures a low splice loss (~0.2 dB) can be obtained by trained field installers. Each cable has a total of 4x NANF splices and 2x NANF-SSMF adapters (insertion loss ~0.5 dB), to provide a practical and flexible way of connecting the cable to SSMF components in the lab (i.e., to terminal and diagnostic equipment). This configuration is representative of a real-world cable deployment. The total insertion loss of the 10.15 km riser cable including splices and adapters was 10.0 dB and the 10.35 km LT cable loss was 9.9 dB. Note that the isolator is inserted between the cables to limit any potential reflections from the adapters. This would not be present in a field-deployment since the cables would be directly spliced together.

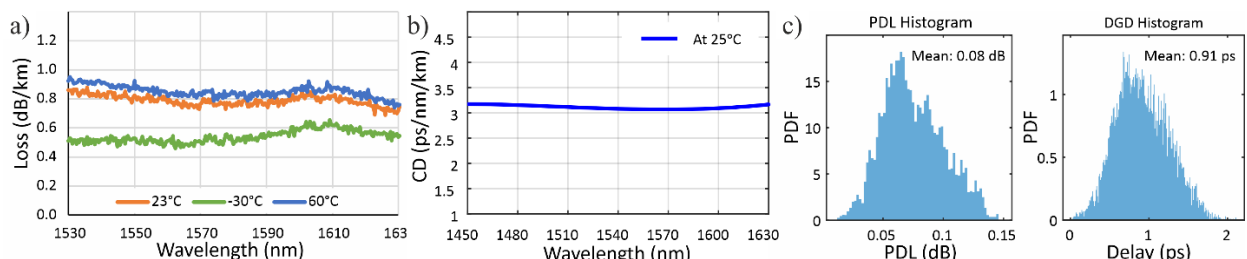

Figure 2: (a) Spectral loss over temperature and (b) Chromatic Dispersion (CD) of NANF LT Cable; (c) Polarization Dependent Loss (PDL) and Differential Group Delay (DGD) distributions measured on the ~10 km concatenated LT transmission line.

Key optical properties of the NANF LT cable are shown in Figs 2.a-c., with very similar performance achieved in the riser cable. Fig. 2a, the cable loss is < 0.95 dB/km over a -30°C to +60°C temperature range. Fig. 2b shows the chromatic dispersion of our NANF cable, which varies between just 3 and 3.25 ps/nm.km over the range 1450 to 1630 nm, far lower and flatter than those of SSMF. Moreover, the NANF cable has effectively no optical nonlinearity. Finally, the histograms in Fig. 2c illustrate the NANF LT cable DGD and PDL, both of which confirm very good cabled performance. Note the substantially reduced loss (by a factor of 5-10), increased bandwidth (>100 nm vs 18.7 nm) and an order of magnitude lower CD and DGD versus those previously reported for PBGF HCF cable [6].

### 3. Transmission Experiments with NANF Cable

Fig. 3a depicts the link used to demonstrate a field-deployable NANF cable with commercial equipment in a high throughput DCI application. It comprises Ciena line equipment (EDFAs, WDM multiplexer (MUX) and demultiplexer (DEMUX)) as well as 20.5 km of cabled hollow-core NANF (see Fig. 1c). An optical isolator disables the post amplifier return loss shutoff. Two types of transmission equipment, namely WaveLogic 5 Extreme line rate adjustable 200G-800G coherent optics and WaveLogic 5 Nano 400ZR QSFP-DD coherent pluggable, WL5e and WL5n, respectively, are tested separately. In each test a bulk modulated WDM spectrum (47 WL5e channels or 63 WL5n channels) is combined with a MUX-filtered probe channel and measurements are made for each channel location in the C-band by tuning the channel under test. Probe transmitter and receiver are provided by separate transceivers. Thus, two WL5e transceivers operating at 91.6 GBd and 800 Gb/s throughput are used for one test, and two WL5n transceivers operating at 59.84375 GBd and 400 Gb/s throughput are used for a second test. Aggregate C-band throughput is 38.4 Tb/s for WL5e and 25.6 Tb/s for WL5n. Bulk modulation is independent for odd and even channels and conforms to the modulation format of the probe channel. Each bulk channel is further decorrelated from the rest with its own length of fiber.

All channels are amplified with a Ciena EDFA and propagated through 20.5 km of cabled NANF. The cables are configured as described in section 2 and illustrated in Fig. 1. The signal at the receive end is amplified, demultiplexed and sent to the receiver. The bit error rate (BER) at the receiver is averaged for 300 seconds, then converted to an effective signal to noise ratio (ESNR): the SNR is obtained from the BER assuming AWGN. Fig. 4 shows the ESNR margin (difference between the measured ESNR and the ESNR at forward error correction (FEC) failure), across the C-band for both WL5e and WL5n transceivers. An aggregate throughput of 38.4 Tb/s is obtained with at least 2.0 dB of ESNR margin for WL5e at 800 Gb/s. Using a WL5n operating at 59.84375 GBd-400 Gb/s, an aggregate 25.6 Tb/s of client traffic throughput is achieved with at least 2.1 dB of ESNR margin. A Viavi ONT-804 Optical Network Tester was used to monitor the client traffic for the channel with least margin. The client traffic was observed for 18 hours and ran error free for both WL5e at 800 Gb/s and WL5n at 400 Gb/s.

The WDM performance of the 20.5 km cabled NANF was also measured in the re-circulating loop of Fig. 3b. For this experiment the transmitter under test is a commercial WL5e transceiver operating at 95 GBd. WL5e traffic wave-forms, at 800 Gb/s or 400 Gb/s, are uploaded as digital DAC instructions to a transmitter memory. For WDM measurements, the test channel is combined with 44 like traffic bulk modulated WDM channels. All channels are amplified by a Ciena EDFA before entering the re-circulating loop. This last contains 20.5 km of cabled NANF, a synchronous polarization scrambler (PS), a wavelength selective switch (WSS) for gain flattening and a variable optical attenuator (VOA) for loop power balancing. The PS applies a random polarization rotation to each loop orbit. Three Ciena EDFAs compensate the total loop loss. The signal exiting the loop is amplified and shared between the OSA and coherent receiver (CoRx). The 70.5 µs loop latency did not allow use of a WL5e receiver in these experiments. Instead, the signal is detected with a WL5e optical receiver module and a Keysight UXR-Series oscilloscope converts the analog signal into digital to be processed offline using a floating-point version of WL5e ASIC DSP. The resulting transceiver is measured to be -25 dB noisier than the WL5e product. The maximum reach at zero ESNR margin is measured for five test channels spanning the C-band. The results are summarized in Fig. 5. A minimum propagation distance of 1128 km at 400 Gb/s and 126 km at 800 Gb/s was achieved. We measured spectral variation in reach (2000 km at 192.6 THz vs. 1100 km at 195.8 THz). This is a legacy artifact of the manufacturing process, is not fundamental to the design and we are confident that all channels will achieve maximum reach in future cable generations.

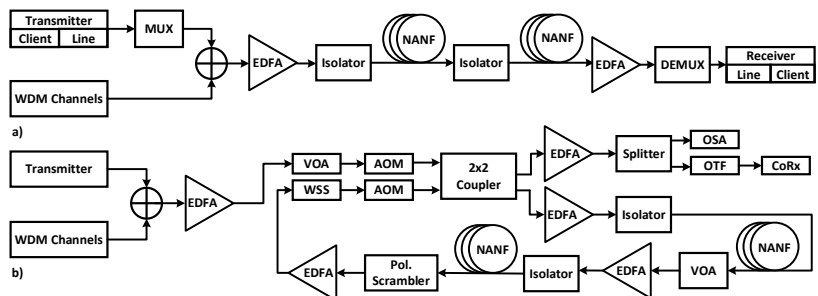

Figure 3: Experimental setups: (a) point to point commercial transmission system and (b) re-circulating loop for reach investigation. All amplifiers operate at a total output power of 23.5 dBm.

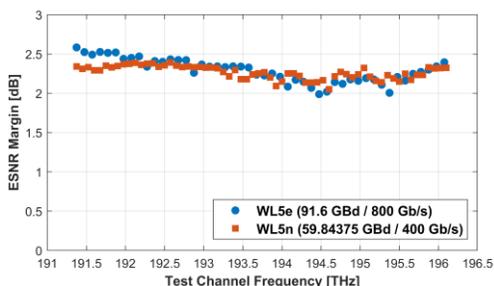

Figure 4: ESNR margin vs. channel frequency over commercial WDM transmission system for 48 WL5e channels operating at 800 Gb/s and 64 WL5n channels operating at 400 Gb/s.

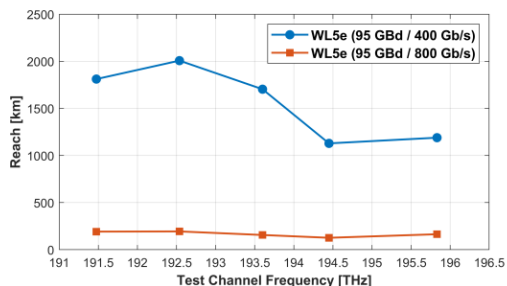

Figure 5: Reach vs. frequency for 45 WDM channels propagating in re-circulating loop for WL5e operating at 95 GBd 800 Gb/s and 400 Gb/s.

## 4. Conclusion

We have reported a breakthrough in field-deployable cabled NANF hollow-core fiber transmission, demonstrating 38.4 Tb/s (800Gb/s-x-48 WDM-channels) over 20.5 km using commercial terminal equipment. This distance was limited by the length of cable in the experiment. We have also shown 400/800Gb/s WDM transmission over distances between 1128/126 km and 2007/194 km, depending on C-band frequency. These results exceed previously reported cabled HCF performance by orders-of-magnitude in terms of reach and throughput and shows the potential for the use of NANF hollow-core cable in future metro and long-haul systems.

## 5. References

[1] F. Poletti et al., "Towards high-capacity fibre-optic communications at the speed of light in vacuum," Nature Photonics,7 pp. 279–284, 2013.
[2] A. Osipovich, "High-Frequency Traders Push Closer to Light Speed With Cutting-Edge Cables", Wall Street Journal, December 15 2020.
[3] B. J. Mangan et al., "Low loss (1.7 dB/km) hollow core photonic bandgap fiber," in OFC, PDP 24, 2004.
[4] F. Poletti, "Nested antiresonant nodeless hollow core fiber," J. Opt. Express, vol. 22, no. 20, pp. 23807–23828, 2014.
[5] G. T. Jasion et al., "Hollow Core NANF with 0.28 dB / km Attenuation in the C and L Bands," in OFC Th4B.4, 2020.

[6] A. Nespola et al. "Transmission of 61 C-band channels over record distance of hollow core fiber with L-band interferers", IEEE J. Lightwave. Tech. vol 29, pp. 813-820, 2021.
[7] B. Zhu et al., "First demonstration of Hollow Core Fiber Cable for Low Latency Data Transmission," in OFC, Th4B.3, 2020.